\documentclass[runningheads, 10pt]{llncs}
\usepackage{orcidlink}
\usepackage[letterpaper, margin=1in]{geometry}
\usepackage{comment}
\usepackage[utf8]{inputenc}
\usepackage{multirow}
\usepackage{pdflscape} 
\usepackage{booktabs} 
\usepackage{soul,color}

\usepackage[T1]{fontenc}
\usepackage{graphicx}
\usepackage{subcaption}
\def\BibTeX{{\rm B\kern-.05em{\sc i\kern-.025em b}\kern-.08em
    T\kern-.1667em\lower.7ex\hbox{E}\kern-.125emX}}
\begin{document}

\title{HiCat: A Semi-Supervised Approach for Cell Type Annotation}

\titlerunning{HiCat: A Semi-supervised Cell Annotation Pipeline}
%
\author{Chang Bi\inst{1}\orcidlink{0009-0003-2660-2018}, Kailun Bai\inst{1}, Xing Li\inst{2}$^{,*}$\orcidlink{0000-0002-4186-7909}, Xuekui Zhang\inst{1}$^{,*}$\orcidlink{0000-0003-4728-2343}}
\authorrunning{Bi et al.}
%
\institute{
University of Victoria, Victoria BC V8W 2Y2, Canada\\
\and
University of Saskatchewan, Saskatoon SK S7N 5A2, Canada\\
$^{*}$ Corresponding authors: \email{xuekui@uvic.ca, lix491@usask.ca}
}
\maketitle              
\begin{abstract}
We introduce HiCat (\underline{H}ybr\underline{i}d \underline{C}ell \underline{A}nnotation using \underline{T}ransformative embeddings), a novel semi-supervised pipeline for annotating cell types from single-cell RNA sequencing data. HiCat fuses the strengths of supervised learning for known cell types with unsupervised learning to identify novel types. This hybrid approach incorporates both reference and query genomic data for feature engineering, enhancing the embedding learning process, increasing the effective sample size for unsupervised techniques, and improving the transferability of the supervised model trained on reference data when applied to query datasets. The pipeline follows six key steps: (1) removing batch effects using Harmony to generate a 50-dimensional principal component embedding; (2) applying UMAP for dimensionality reduction to two dimensions to capture crucial data patterns; (3) conducting unsupervised clustering of cells with DBSCAN, yielding a one-dimensional cluster membership vector; (4) merging the multi-resolution results of the previous steps into a 53-dimensional feature space that encompasses both reference and query data; (5) training a CatBoost model on the reference dataset to predict cell types in the query dataset; and (6) resolving inconsistencies between the supervised predictions and unsupervised cluster labels. When benchmarked on 10 publicly available genomic datasets, HiCat surpasses other methods, particularly in differentiating and identifying multiple new cell types. Its capacity to accurately classify novel cell types showcases its robustness and adaptability within intricate biological datasets. Except for this sentence, this document is exactly the same as the submitted version for RECOMB 2025 on October 28, 03:06 GMT.

Code is available on Github: \url{https://github.com/changbiHub/HiCat}

\end{abstract}

\newpage
\section{Introduction}

Single-cell RNA sequencing (scRNA-seq) has revolutionized bioinformatics and biomedicine by providing unprecedented insights into cellular heterogeneity, rare cell types, and cell-to-cell interactions \cite{jovic_singlecell_2022,bridges_mapping_2022}. This technology is pivotal for advancing our understanding of complex biological systems and diseases. In cancer research, for example, scRNA-seq enables researchers to dissect tumor heterogeneity, offering a clearer picture of how diverse cell populations contribute to disease progression and treatment resistance \cite{sallese_abstract_2024,zhang_mast_2023}. Its applications extend to developmental biology, where scRNA-seq helps elucidate the dynamic processes of cell differentiation, and immunology, by revealing the intricate landscape of immune cell diversity and function \cite{bridges_mapping_2022}. These insights are crucial for developing targeted therapies and personalized medicine approaches.

Unsupervised machine learning, particularly through clustering analysis, is a prominent method for annotating cell types. These techniques cluster cells using a gene-by-cell expression matrix and subsequently label each resulting cluster. Despite their effectiveness and popularity, these methods encounter two significant limitations. First, clustering annotations often depend on feature genes from the literature, making the process labor-intensive and heavily reliant on expert input. This dependency raises issues with scalability and reproducibility \cite{pasquini_automated_2021,liu_exploring_2023}. Recently, anchor-based annotation approach has recently emerged as a partial solution, exemplified by Seurat version 3 or higher version \cite{stuart_comprehensive_2019}, leveraging another single-cell RNA-seq dataset (known as the reference set) that includes annotated cell type labels. Cells in the reference set are grouped by their cell types, and each cluster is anchored to specific cell type groups in the reference set and labeled accordingly. Second, unsupervised annotation is often prone to errors from cluster impurity, leading to misannotations \cite{pasquini_automated_2021}, an issue that lacks a satisfactory remedy. Such challenges can hinder the discovery of new cell types and complicate our understanding of biological processes. With the increasing amount of scRNA-seq data, there is a critical need for enhanced annotation methods that are both efficient and accurate \cite{liu_exploring_2023}.

With the growing use of single-cell sequencing technology, an increasing amount of annotated single-cell genomic data has become available to researchers. This trend has driven the development of supervised learning techniques for cell type annotation in single-cell genomics. These methods leverage labeled reference datasets to classify individual cells in query data, thereby addressing issues of cluster impurity by annotating cells rather than entire clusters. Various general-purpose supervised machine learning models have been employed to create cell annotators. For instance, Garnett\cite{pliner_supervised_2019} use Elastic Net for cell type classification. CaSTLe\cite{lieberman_castle_2018} uses XGBoost\cite{chen_xgboost_2016}, and SingleCellNet\cite{tan_singlecellnet_2019} applies random forests\cite{breiman_random_2001} for cell classification based on selected genes and derived features. scPred\cite{alquicira-hernandez_scpred_2019} and Moana\cite{wagner_moana_2018} utilize support vector machines (SVM)\cite{saunders_support_1998} for classification. Correlation-based approaches have also been widely used for cell annotation, as seen in methods such as SingleR\cite{aran_reference-based_2019}, scmap-cluster\cite{kiselev_scmap_2018}, CHETAH\cite{dekanter_chetah_2019}, scMatch\cite{hou_scmatch_2019}, ClustifyR\cite{fu_clustifyr_2020}, and CIPR\cite{ekiz_cipr_2020}. K-nearest neighbor classifiers\cite{cover_nearest_1967} form part of the pipelines in scmap-cell\cite{kiselev_scmap_2018}, Moana\cite{wagner_moana_2018}, OnClass\cite{noauthor_leveraging_nodate}, scANVI\cite{xu_probabilistic_2021}, and scClassify\cite{lin_scclassify_2020}. scID\cite{boufea_scid_2020} employs linear discriminant analysis for cell type annotation, while scAnnotate\cite{ji_scannotate_2023} incorporates generative learning and ensemble techniques. To mitigate batch effects between reference and query datasets, scMagic\cite{noauthor_scmagic_nodate} performs a second round of reference-based classification. In addition to traditional machine learning and statistical methods, deep learning approaches have also been developed for cell annotation, including SuperCT\cite{xie_superct_2019}, ACTINN\cite{ma_actinn_2020}, scDeepSort\cite{shao_scdeepsort_2021}, and EpiAnno\cite{chen_cell_2022}.

Supervised cell annotation methods classify cells based on known cell types present in the reference set, making them unable to identify truly novel cell types. Some methods, such as Scmap\cite{kiselev_scmap_2018}, CHETAH\cite{dekanter_chetah_2019}, SingleCellNet\cite{tan_singlecellnet_2019}, scClassify\cite{lin_scclassify_2020}, and scID\cite{boufea_scid_2020}, can assign an "unassigned" label when a cell is considered too dissimilar to any known types. While this feature can help in detecting novel cell types, it is not ideal because these methods optimize feature selection primarily for distinguishing between known categories. In real data experiments, we observed a decline in annotation performance as the proportion of unknown cell types increased. Additionally, these methods are unable to differentiate between multiple distinct unseen cell types, a capability that is a strength of unsupervised approaches. This suggests that using semi-supervised learning methods, which combine the benefits of both supervised and unsupervised approaches, may provide a more effective solution.

Semi-supervised learning combines the strengths of supervised and unsupervised learning to improve cell type annotation accuracy. It leverages labelled reference data while incorporating information from unlabeled data, resulting in more robust predictions even when labelled data is limited. This is particularly valuable in scRNA-seq contexts where cell type diversity is vast, and labeled data availability is limited \cite{liu_exploring_2023}. Recently semi-supervised methods were developed for cell annotation. For example, CALLR\cite{wei_callr_2021} utilizes unlabeled data by combining a spares logistic regression and a Laplacian matrix constructed from all cell for clustering. scNym\cite{kimmel_scnym_2020} adapts pseudo-labeling technique and Domain Adversarial Networks(DAN)\cite{ganin_domain-adversarial_2016} to incorporates unlabeled data into the training process. scSemiGan\cite{xu_scsemigan_2022} uses a modified Generative Adversarial Networks\cite{goodfellow_generative_2014}, that uses labeled data to guide the training process, while leverages unlabeled data by generating latent representations that capture underlying structures. scBERT\cite{yang_scbert_2022} performs a self-supervised pre-training on unlabeled data then fine-tuning the model on labeled data.

Despite the availability of multiple semi-supervised methods, most only focus on classifying known cell types. None, to our knowledge, explicitly discuss how to distinguish between novel cell types or implement this in their software. This motivated us to utilize a semi-supervised approach to discover novel cell types and distinguish different types of novel cells. Furthermore, all current methods work on feature space with single resolution (i.e. either work with genes or work with dimension-reduced features). While dimension reduction methods generate results in multiple resolutions, and all have their specific strength, utilizing information from features at different resolutions is desired. This study introduces a novel semi-supervised pipline designed to address these challenges using a semi-supervised learning approach. 

To address the abovementioned issues, we present HiCat, a novel cell annotation pipeline based on semi-supervised learning, with four key advantages. First, HiCat uses sequential dimension reduction to create a multi-resolution feature space from both reference and query genomic data, embedding both into a unified space that makes the model more transferable for annotating query cells. Second, CatBoost is trained on this unified feature space as a supervised model composed of numerous shallow trees that sequentially address misclassified samples. Each tree automatically selects the most relevant multi-resolution features for its specific task, fully leveraging the multi-resolution space. Third, HiCat provides unsupervised labels using DBSCAN, which excels at detecting small, unknown clusters and distinguishing rare cell types within the query set. Finally, HiCat’s final prediction combines supervised labels from CatBoost and unsupervised labels from DBSCAN, retaining the ability to identify unknown cell types while minimizing cluster impurity issues common in unsupervised methods. Comprehensive benchmarking with real single-cell RNA-seq datasets shows HiCat’s competitive performance in classifying known cell types and superior ability to identify unknown cell types, including rare ones with as few as 20 cells in the query data, and its unique feature of successfully differentiating multiple unknown cell types.

\section{Method}

\begin{figure}[htbp]
\centerline{\includegraphics[width=0.8\textwidth]{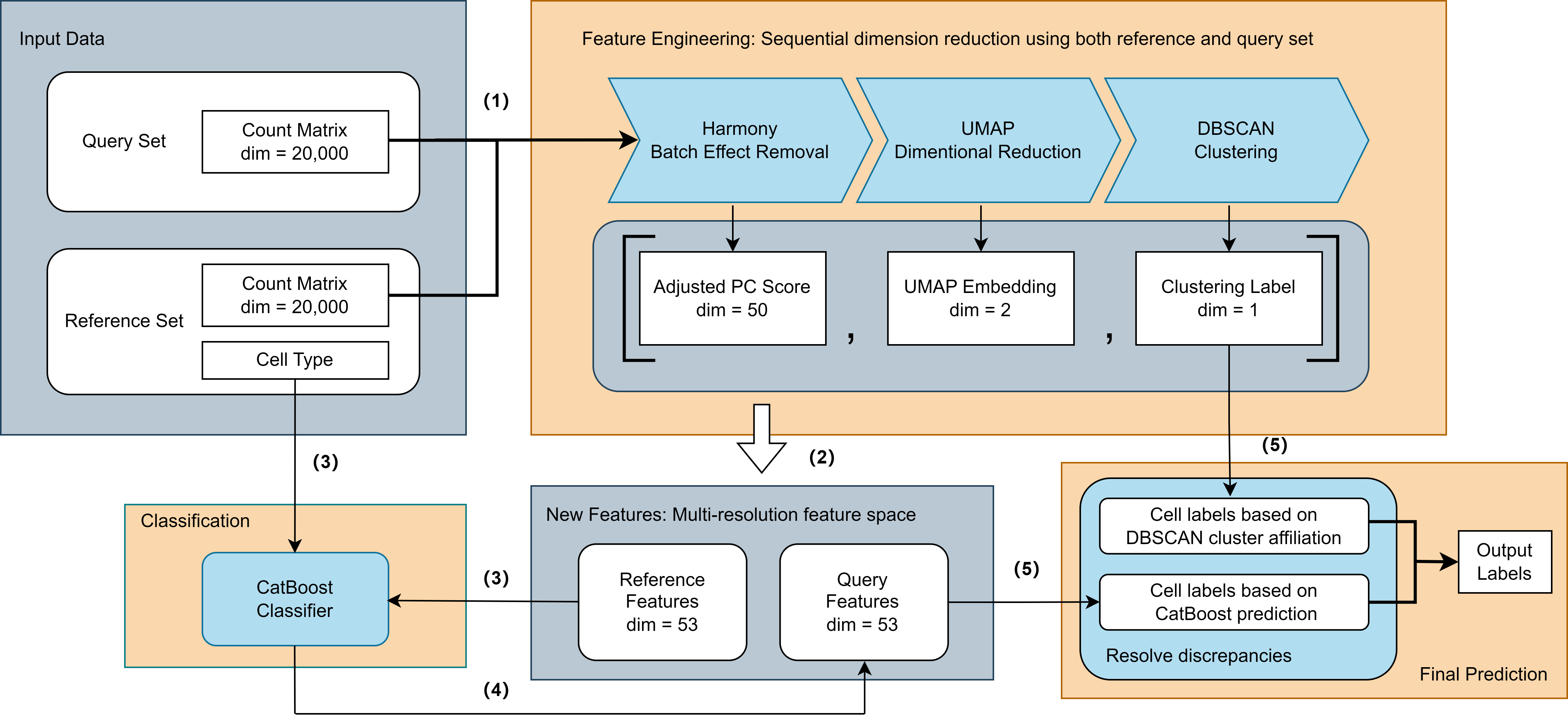}}
\caption{Workflow of HiCat. \textbf{(1) Sequential Dimension Reduction:} Harmony is applied to remove batch effects between the genomic matrices of the reference and query sets, reducing the data to 50 principal components (PCs). The data is then further reduced by UMAP and clustered using DBSCAN. \textbf{(2) Concatenation of Multi-Resolution Features:} The aligned PC scores, UMAP embeddings, and cluster labels are combined to create a 53-dimensional multi-resolution feature space derived from both reference and query datasets. \textbf{(3) Model Training:} The 53-dimensional features along with the cell type labels from the reference set are used to train a CatBoost classifier. \textbf{(4) Classification of Query Cells:} The trained classifier is applied to the new features of the query set, providing cell type annotations with predictive probabilities as indicators of confidence. \textbf{(5) Resolving Discrepancies:} Discrepancies between the DBSCAN and CatBoost outputs are resolved to determine the final cell type labels; the CatBoost prediction is used when its predictive probability is high, otherwise, the DBSCAN label is assigned.}
\label{fig:diagram}
\end{figure}

The input data of cell annotation consists of two genomic matrices: one from the reference set and the other from the query dataset. In these matrices, the rows represent tens of thousands of genes, while the columns indicate cells, typically numbering in the hundreds of thousands. The input also includes cell type labels for each cell in the reference set, but the query data does not have these labels. The main goal of the cell annotation analysis is to predict these absent labels.

HiCat (Hybrid Cell Annotation using Transformative embeddings) is a semi-supervised framework created for annotating cell types in single-cell RNA sequencing (scRNA-seq) data. Figure~\ref{fig:diagram} illustrates the different components of the HiCat pipeline. In the rest of Method section, we present a comprehensive overview of these components in six analysis steps, concluding with a summary of the method's innovative aspects.

\paragraph{\textbf{Step 1. Batch Effect Removal Using Harmony}}: To align the reference and query datasets, we start by using Harmony \cite{korsunsky_fast_2019}, an algorithm developed to correct batch effects in multi-dataset integrations. We choose Harmony since it was recommended as top 3 batch effect removal methods in a recent benchmark study \cite{tran_benchmark_2020} Harmony functions within the principal components (PCs) space, iteratively adjusting the data to synchronize shared cell types across datasets, all while preserving biological variation. In our pipeline, we implement Harmony on the top 50 principal components, its default configuration, for both the reference and query datasets. This ensures that the shared patterns are retained as batch-specific variation is eliminated. The outcome is a harmonized 50-dimensional embedding for both datasets. We input the reference and query datasets containing tens of thousands of genes, and the result is a 50-dimensional principal component embedding that forms the basis for further analysis.

\paragraph{\textbf{Step 2. Dimensionality Reduction Using UMAP}}: Following Harmony-based alignment, we further reduce the data's dimensionality using Uniform Manifold Approximation and Projection (UMAP) \cite{mcinnes_umap_2020}. We opt for UMAP, the leading method for reducing dimensions in visualizing single-cell genomic data. It is a non-linear technique for dimensionality reduction that effectively captures both local and global structures within high-dimensional datasets. We apply UMAP to the 50-dimensional embedding obtained in the prior step, compressing it down to 2 dimensions to reveal the two most critical patterns in the data. The input for this process is the 50-dimensional embedding produced by Harmony, while the output is a 2-dimensional UMAP embedding that highlights the most critical data patterns.

\paragraph{\textbf{Step 3. Clustering with DBSCAN}}: Next, we conduct unsupervised clustering using Density-Based Spatial Clustering of Applications with Noise (DBSCAN) \cite{ester_density-based_1996}. We use DBSCAN over other clustering methods, since it offers several key advantages, particularly in the context of single-cell genomic data. Unlike K-means, DBSCAN does not require the number of clusters to be predefined and can detect clusters of varying shapes and densities, making it highly flexible. It excels at identifying outliers or noise, which is particularly useful for recognizing low-quality cells or rare populations that might otherwise be obscured. Unlike hierarchical clustering, which assigns all data points to clusters, DBSCAN can leave some cells unassigned, highlighting rare or ambiguous cases. Additionally, DBSCAN works well when paired with dimensionality reduction techniques like UMAP or PCA, effectively clustering lower-dimensional embeddings. These advantages make DBSCAN especially useful for the complex, high-dimensional structure of single-cell RNA-seq data, where cluster shapes, densities, and outliers vary significantly. DBSCAN produces a one-dimensional vector where each cell is assigned a cluster label, or marked as noise if it doesn't fit into any cluster. The input for DBSCAN consists of the 2-dimensional UMAP embeddings, while the output is a one-dimensional vector reflecting each cell’s cluster affiliation.

\paragraph{\textbf{Step 4. Feature Space Concatenation}}: To fully integrate biological and structural insights from both datasets, we combine the results from steps 1 to 3. This process merges the 50-dimensional principal component embedding from Harmony, the 2-dimensional UMAP embedding, and the one-dimensional cluster membership vector from DBSCAN. The result is a 53-dimensional feature space representing various resolution levels, encompassing known and unknown patterns from the reference and query datasets. The inputs for this phase include the outputs from Harmony, UMAP, and DBSCAN, which together form this 53-dimensional feature space for the genomic data of both reference and query datasets.

\paragraph{\textbf{Step 5. Supervised Learning Using CatBoost}}: For supervised cell types annotation, we train a CatBoost classifier \cite{dorogush_catboost_2018} with the reference dataset and its corresponding cell type labels. CatBoost is a gradient boosting algorithm that performs exceptionally well with categorical features and includes built-in regularization to prevent overfitting. Using the 53-dimensional feature vectors from the reference dataset, the model is trained to predict cell type labels using the extracted features. Once trained, the model is used to predict the cell type for each cell in the query dataset, along with providing a probability distribution across potential cell types. For this step, the inputs are the 53-dimensional feature vectors and reference cell type labels, while the output is a trained CatBoost model and a probability distribution of cell types for each cell in the query set.

\paragraph{\textbf{Step 6. Resolving Discrepancies Between Supervised and Clustering-Based Labels}}: In Step 3, each cell is annotated based on its cluster affiliation, while in Step 5, cell types are assigned based on the highest predicted probability from CatBoost. Catboost and DBSCAN work integrally to reinforce each other. The cluster affiliations serve as a strong guide for the classifier, improving the accuracy when predicting known cell types; the classifier discriminates low confidence prediction, allowing the cluster affiliation to only apply to cells with potentially unseen types in the final prediction. The predicted probability from CatBoost reflects its confidence in classification. If this probability is close to “1/number of cell types,” it indicates low confidence in CatBoost’s prediction, and we defer to DBSCAN’s cluster affiliation. Conversely, if the predicted probability is close to 1, it signifies high confidence in CatBoost’s result, which should be used. The threshold for determining high confidence is based on the largest drop in predicted probabilities across both the reference and query datasets. The input to this step includes the CatBoost predictions and the DBSCAN cluster labels, and the final output is the annotated cell type labels for the query dataset.

\paragraph{\textbf{Summary of Novelties and Advantages}}: The HiCat pipeline features two novelties contributing to its outstanding performance: (1) Feature engineering and (2) integrating supervised and unsupervised outcomes. Feature engineering involves a sequential dimension reduction process (Steps 1 to 3), which creates a feature space with various resolutions, allowing the classifier in Step 5 to learn information from features at different resolutions. Additionally, feature engineering using genomic data from both reference and query datasets enhances the embedding quality and effectively expands the sample size for unsupervised learning. This also improves the transferability of the supervised model trained on reference data when applied to query data. The pipeline efficiently fuses unsupervised and supervised results based on their confidence levels, leveraging the advantages of both approaches. HiCat accurately labels known cell types when supervised learning generates high-confidence predictions, addressing common issues of cluster impurity found in unsupervised methods. Furthermore, it assigns labels to unknown cells based on clustering outcomes when supervised methods show low confidence in out-of-distribution (OOD) samples, whose cell types are absent from the reference data, enabling the identification and annotation of novel cell populations.

\section{Performance Evaluation}

To evaluate HiCat's performance, we compare it with well-established methods, including Seurat \cite{hao_dictionary_2024}, SingleCellNet \cite{tan_singlecellnet_2019}, scClassify \cite{lin_scclassify_2020}, SingleR \cite{aran_reference-based_2019}, CaSTLe \cite{lieberman_castle_2018}, SCINA \cite{zhang_scina_2019}, scID \cite{boufea_scid_2020}, CHETAH \cite{dekanter_chetah_2019}, and scmap \cite{kiselev_scmap_2018}. Our assessment consists of three experiments: (1) Measuring the annotation performance of all methods when the query data has no unseen cell types. (2) Analyzing their annotation performance with query sets that contain unseen cell types; here, SingleCellNet \cite{tan_singlecellnet_2019}, scClassify \cite{lin_scclassify_2020}, scID \cite{boufea_scid_2020}, SCINA \cite{zhang_scina_2019} and CHETAH \cite{dekanter_chetah_2019} are evaluated alongside HiCat. (3) Evaluating performance in distinguishing multiple unknown cell types. Notably, the mentioned methods do not distinguish between cell types in unknown classes, highlighting a key feature of HiCat. Consequently, we will specifically showcase HiCat's outcomes in this experiment. All methods are used with their default settings to prevent bias from our ability to fine-tune different methods.

We utilize 10 published single-cell genomic datasets that often serve as benchmarks for other cell annotation methods. These datasets represent a range of mouse and human tissues and were sequenced via different protocols. They consist of datasets from the mouse pancreas \cite{baron_single-cell_2016}, the human pancreas \cite{baron_single-cell_2016,muraro_single-cell_2016,xin_rna_2016,wang_single-cell_2016,segerstolpe_single-cell_2016,lawlor_single-cell_2017}, Tabula Muris (TM) \cite{tabula_muris_consortium_single-cell_2018}, PBMC \cite{ding_systematic_2020}, and the mouse brain \cite{tasic_adult_2016,campbell_molecular_2017}. From these datasets, we derive 33 independent pairs of reference and query sets for external validation. These pairs are structured to ensure the two sets are comparable, e.g., both from the pancreas or both from PBMC, but not a pancreas reference paired with a PBMC query. Supplementary Table outlines the detailed information of these 33 pairs from the datasets used in our first two experiments. The cell type labels of the query sets serve as the ground truth for assessing the classification performance of the predicted labels from the annotation methods. Further details on dataset acquisition and preprocessing can be found in Bai et al. \cite{bai_pclda_2023}, as this benchmark closely follows the procedures outlined in that paper. 

\begin{figure}[ht]
\centerline{\includegraphics[width=0.7\textwidth ]{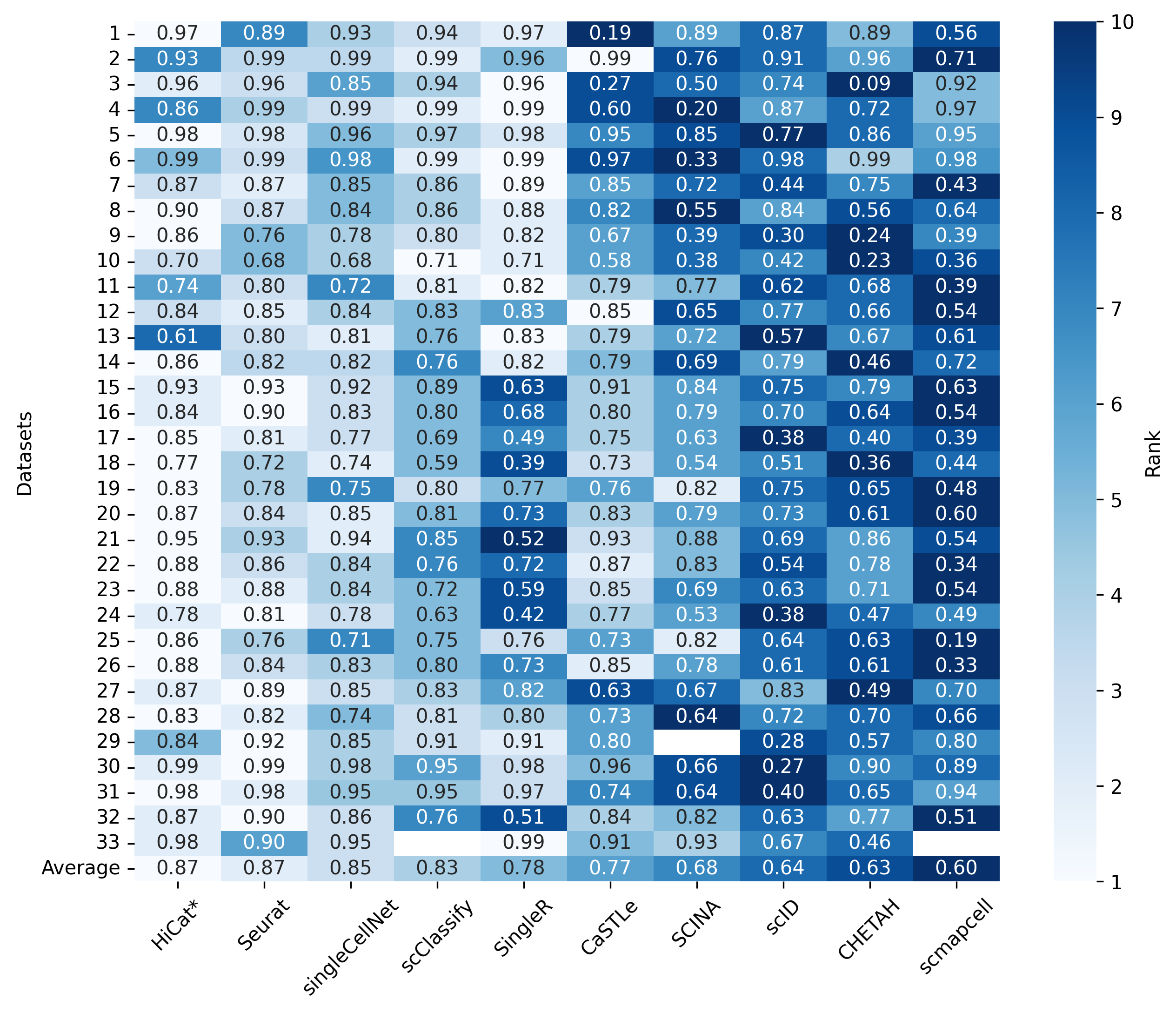}}
\caption{Performance evaluation on data without unseen cell types. Each row depicts a pair of reference and query sets. Columns illustrate the performance of different methods, with each cell's number reflecting the accuracy of the method applied to the corresponding dataset pair. The last row displays the average accuracy for all datasets per method. Color coding indicates the ranking of methods within each dataset, where lighter hues signify superior performance. The findings indicate that HiCat consistently ranks highest across the majority of datasets.}\label{f:seen}
\end{figure}
\paragraph{\textbf{Performance of annotating seen cell types}}: The first experiment assesses how effectively HiCat annotates cells in the query set that match cell types found in the reference set. In each dataset pair, we designate the dataset with a greater variety of cell types as the reference and the other as the query set, ensuring that the query set does not include any unseen cell types. We use the overall accuracy as the evaluation metric for this task, which is defined as the proportion of cells assigned with the correct cell type. Figure~\ref{f:seen} shows HiCat has superior performance compared to other methods in this comparison across most datasets, achieving the highest overall performance among the selected datasets.

\paragraph{\textbf{Performance of annotation when unseen cell types present}}: In the second experiment, we evaluate HiCat’s annotation performance across three scenarios involving unseen cell types: (1) one unseen type, (2) two unseen types, and (3) three unseen types. In each scenario, the difference in the number of cell types between the reference and query datasets remains consistent. To achieve this, we randomly remove cell types to create the desired discrepancies in the number of cell types. Our evaluation focuses on two main aspects of annotation performance. First, we measure the overall accuracy when unseen types are present, calculated similarly to the first experiment but with an additional category labeled "unseen" or "unassigned," representing cells that do not match any cell type in the reference dataset. Second, we evaluate the ability to identify cells as "unseen," treating this as a binary classification problem and utilizing the F1 score, the harmonic mean of precision and recall, a common metric in machine learning for binary classification. Among the methods compared to HiCat in the first experiment, six of them (scmapcell, CHETAH, SCINA, scClassify, singleCellNet, scID) can assign an "unassigned" label for cells that cannot be confidently annotated, which we interpret here as an "unseen" label. HiCat specifically labels cells as "unseen" and, when appropriate, categorizes them into subgroups if multiple unseen cell types are detected. However, for the purpose of this analysis, we simplify by treating all unseen cell types as a single category.

\begin{figure}[htbp]
\centerline{\includegraphics[width=\textwidth]{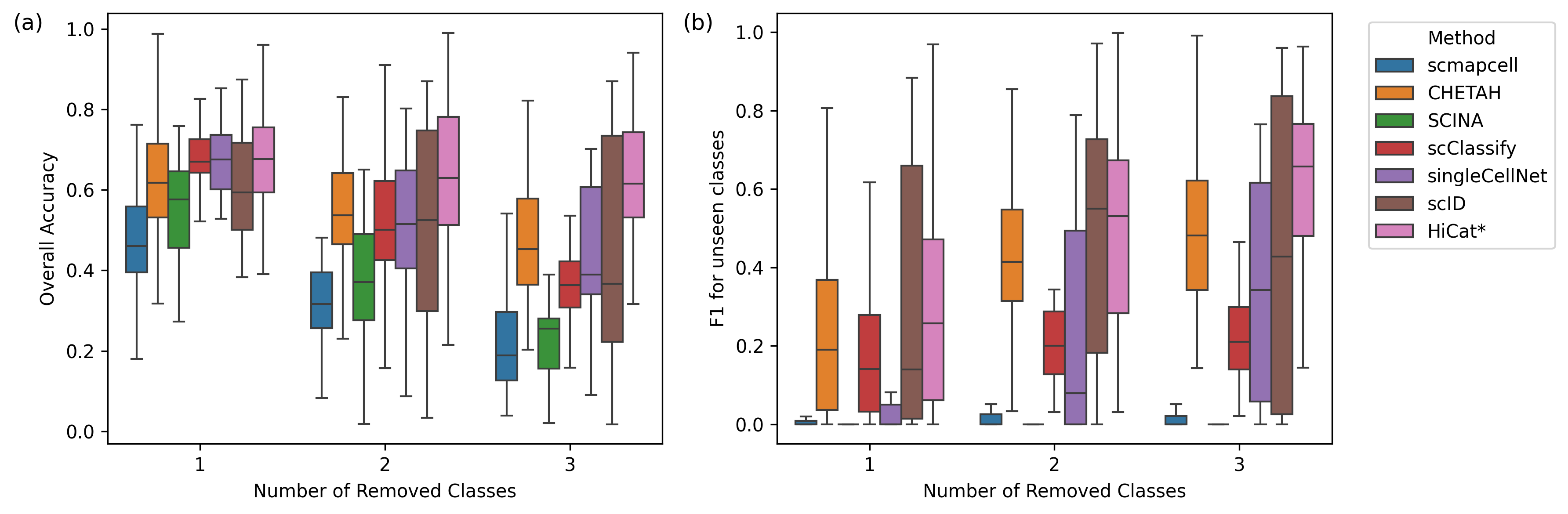}}
\caption{Annotation performance on data with unseen cell types. Each boxplot displays the performance of the respective method across all 33 dataset pairs. The horizontal axis represents the number of unseen cell types in the query set, created by randomly omitting cell types from data. The left panel shows overall accuracy, reflecting each model’s ability to classify all cell types, categorizing missing reference types as 'unseen.' The right panel shows the F1-score for the 'unseen' category, based on precision and recall in identifying these cells. Results indicate that HiCat consistently outperforms other methods when unseen cell types exist.}\label{f:unseen1}
\end{figure}

The left panel of Figure~\ref{f:unseen1} illustrates the overall accuracy of HiCat in comparison to six alternative methods. Each boxplot summarizing the accuracy of a method across 33 data pairs. All six methods demonstrate lower accuracy than in the first experiment, suggesting that the presence of unseen cell types hampers annotation performance universally. HiCat clearly excels, securing the highest overall accuracy. Interestingly, HiCat’s median accuracy remains largely stable despite the increase in unseen cell types, while the accuracy of other methods tends to decline as more unseen cell types are introduced. This implies that the other methods struggle, leading to a higher frequency of incorrect annotations for unseen types, whereas HiCat's robust performance in identifying new cells allows it to sustain relatively steady results.

The right panel of Figure~\ref{f:unseen1} presents the F1 scores for each method's binary classification in identifying unseen cells, with HiCat once again being the leading performer. The relationship between the number of unseen cell types and the F1 score differs from the trend observed for overall accuracy, as the F1 score represents a balanced summary of precision and recall and is influenced by the class distribution. In this experiment, the proportion of unseen cells is relatively small compared to the total number of cells. As more cell types are removed, the balance between seen and unseen cells improves, impacting the F1 scores. 

\paragraph{\textbf{Distinguishing Multiple Unseen Cell Types}} In the third experiment, we demonstrate HiCat’s ability to distinguish between multiple unseen cell types. As this capability is unique to HiCat, no comparison is made with other methods. We illustrate this using examples from two data pairs. The pancreas dataset from Xin et al.\cite{xin_rna_2016} is selected as the reference set because it includes cell types (alpha, beta, delta, and gamma) commonly found in other pancreas datasets. Two pancreas datasets, Muraro et al.\cite{muraro_single-cell_2016} and Segerstolpe et al.\cite{segerstolpe_single-cell_2016}, are used as query sets because they contain all cell types present in the Xin dataset as well as several additional cell types, such as acinar, ductal, stellate, and endothelial.   

\begin{figure}[htbp]
    \centering
    \begin{subfigure}{0.48\textwidth}
        \centering
        \includegraphics[width=\linewidth]{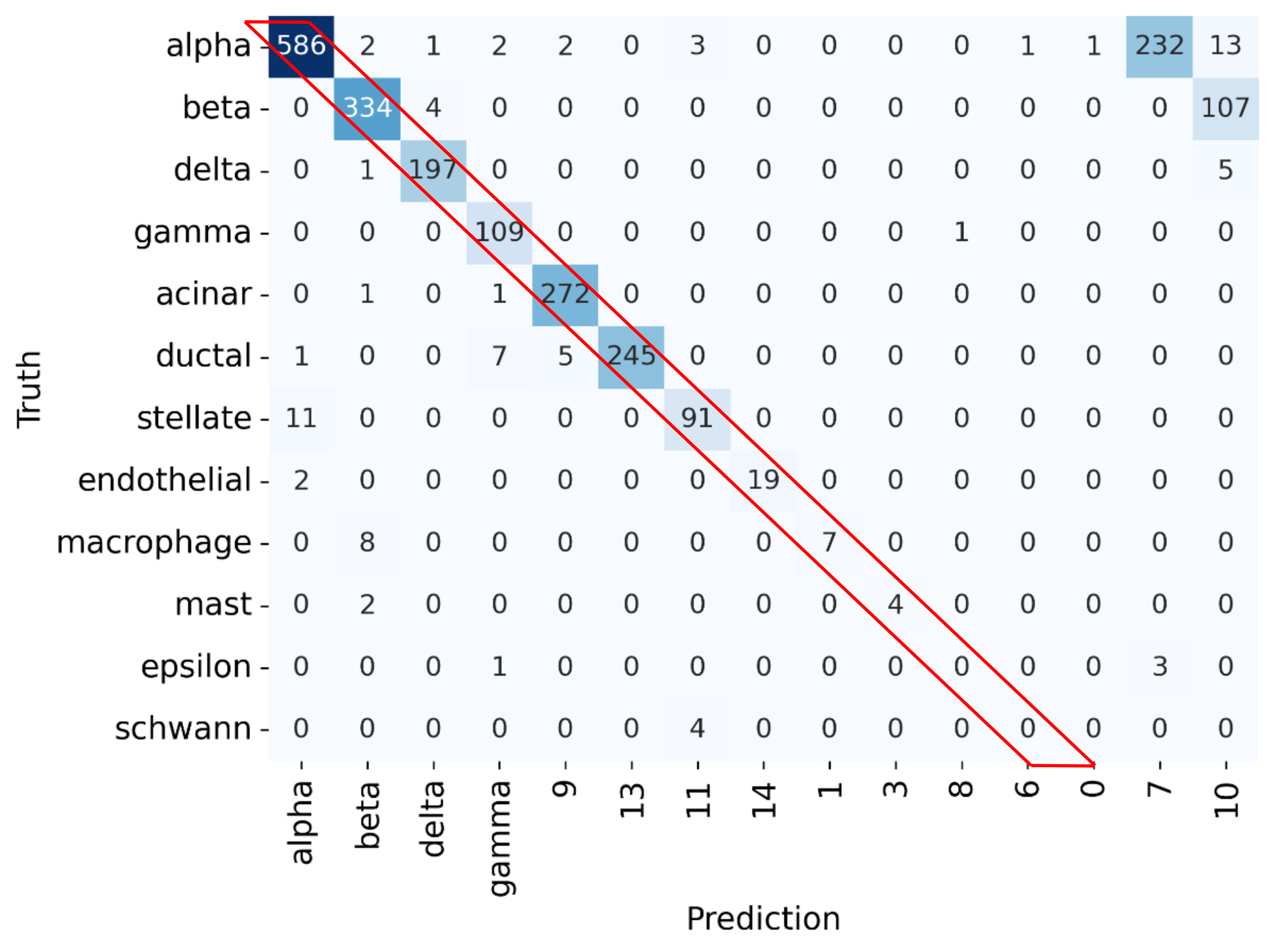}
        \begin{picture}(0,0)
            \put(-110,170){\scriptsize(a)}
        \end{picture}
        \label{fig:xin_Muraro}
    \end{subfigure}%
    \begin{subfigure}{0.48\textwidth}
        \centering
        \includegraphics[width=\linewidth]{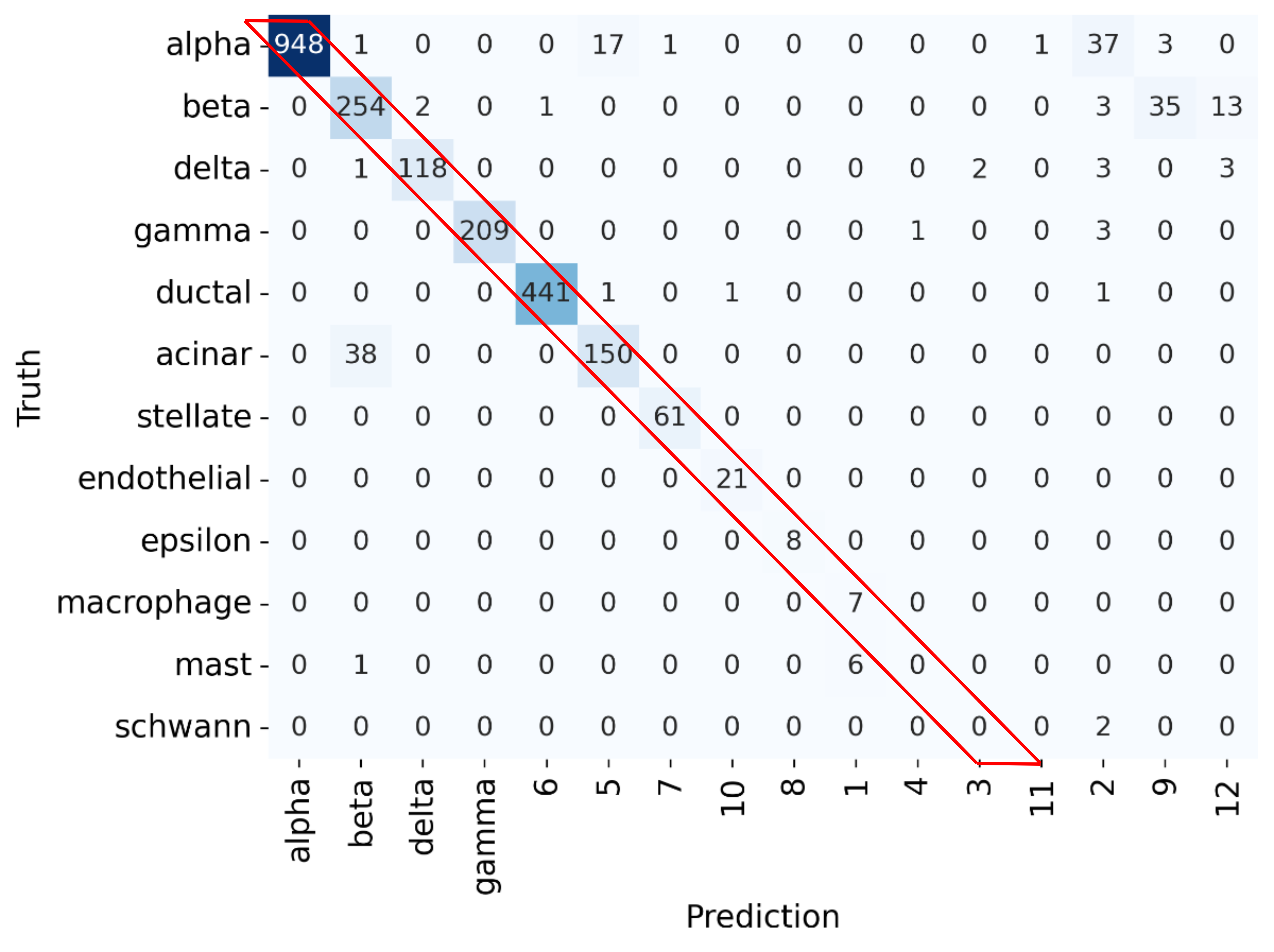}
        \begin{picture}(0,0)
            \put(-110,170){\scriptsize(b)}
        \end{picture}
        \label{fig:xin_seperstophe}
    \end{subfigure}
    \caption{Confusion matrices for cell annotation involving multiple unknown cell types are shown. The rows correspond to the cell labels from the query set, based on published data considered as the ground truth, while the columns represent the cell types predicted by HiCat. Since the reference set contains only four cell types, any additional, unseen types are assigned numerical labels as cell type IDs in the columns. Each element in the confusion matrix indicates the number of cells in the query set with a published label (row) that matches a predicted label (column). The left panel (a) presents the results using Muraro et al. as the query set, while the right panel (b) shows the results using Segerstolpe et al. as the query set. Cluster labels are arranged to align as closely as possible with the likely true labels, with the counts concentrated along the diagonal indicating correct annotations. Leveraging the strength of DBSCAN in detecting small clusters, HiCat successfully distinguished many unknown cell types, even when they contained fewer than 10 cells.}
    \label{fig:combined_figures}
\end{figure}

Figure \ref{fig:combined_figures} presents the confusion matrices for cell annotation on two query datasets. Row names are the labels provided by published data, which are used as ground truth. Column names are HiCat’s predicted labels. The numbers along the diagonal (highlighted by the red frame) represent the count of correctly annotated cells, while the off-diagonal numbers indicate misannotated cells. In this analysis, we focus on HiCat’s performance in distinguishing unseen cell types, which are assigned numerical ID labels for predicted cell types. In both query sets, there are four larger unseen cell types, ranging from 21 to 444 cells, including acinar, ductal, stellate, and endothelial. HiCat accurately annotated all of these cell types, achieving over 90\% accuracy and assigning distinct numerical IDs to each type. For instance, in Figure~\ref{fig:combined_figures}(b), the ductal cells were the best-annotated, with 441 out of 444 cells correctly identified as cluster 6, corresponding to 99.3\% sensitivity. Moreover, only one cell within cluster 6 was not ductal, resulting in a high specificity of 99.8\%. It is impressive to show HiCat's ability to consistently identify the unseen cell type (endothelial) with high accuracy in both query sets, despite having only about 20 cells in both query sets. This success is driven by two key factors: (1) embedding both the reference and query sets in a unified feature space, which allows CatBoost to effectively gauge when its prediction confidence is low, and (2) DBSCAN's strength in detecting smaller clusters, making it particularly well-suited for accurately identifying rare or less abundant cell types. The annotation performance of epsilon cells are not consistent in two query sets. Figures~\ref{fig:combined_figures}(a) has 4 epsilon cells which are not well annotated by HiCat, while Figure \ref{fig:combined_figures}(b) has 8 epsilon cells which perfectly match HiCat's cluster 8 without any error. This indicates HiCat could identify rare unseen cell types with fewer cells, but its performance could be unstable. 

\paragraph{\textbf{HiCat's Errors and Potential Solutions for Improvement}}
We observed HiCat's three types of errors from Figure~\ref{fig:combined_figures}. We discuss these observed errors and suggest potential approaches for addressing them. 

The first type of error involves unknown clusters (seen as columns on the right-hand side of each confusion matrix) that should belong to a known cell type, such as alpha or beta. This issue arises from the heterogeneity within these cell types, where certain subtypes exhibit gene expression profiles that differ significantly from those observed in the reference data. These errors can often be resolved through manual annotation, using feature genes identified in the literature to recognize them as subtypes of known cell types, and subsequently merging them back into the correct categories. Although this type of error largely reduced HiCat's accuracy in our evaluation, it can actually be seen as an advantage, as it allows for the identification of subtypes rather than being merely considered an annotation mistake. This capability can provide valuable insights into the underlying heterogeneity of cell populations.

The second type of error occurs when HiCat fails to separate similar cell types. For instance, macrophages and mast cells are correctly annotated without error in Muraro’s query set, but they are grouped into a single cluster in Segerstolpe’s query set. This reflects the known similarities in gene expression between macrophages and mast cells, which share roles in immune responses, inflammation, and the release of common cytokines, chemokines, and other mediators \cite{de_filippo_mast_2013,noauthor_mast_nodate}. To address this type of error, it is important to identify cell types with highly similar gene expression profiles and conduct downstream analyses to distinguish them further using curated marker genes specific to those cell types.

The third type of error involves identifying noise clusters containing only one or two cells, observed in both query sets. These clusters typically represent outliers from known cell types. While HiCat’s sensitivity to rare and unknown cell types enables it to detect novel patterns, it also makes it prone to recognizing outliers as distinct clusters. Consequently, caution is advised when interpreting clusters with very few cells, as they are more likely to reflect noise rather than meaningful biological variation.

\begin{figure}[h]
\centerline{\includegraphics[width=\textwidth]{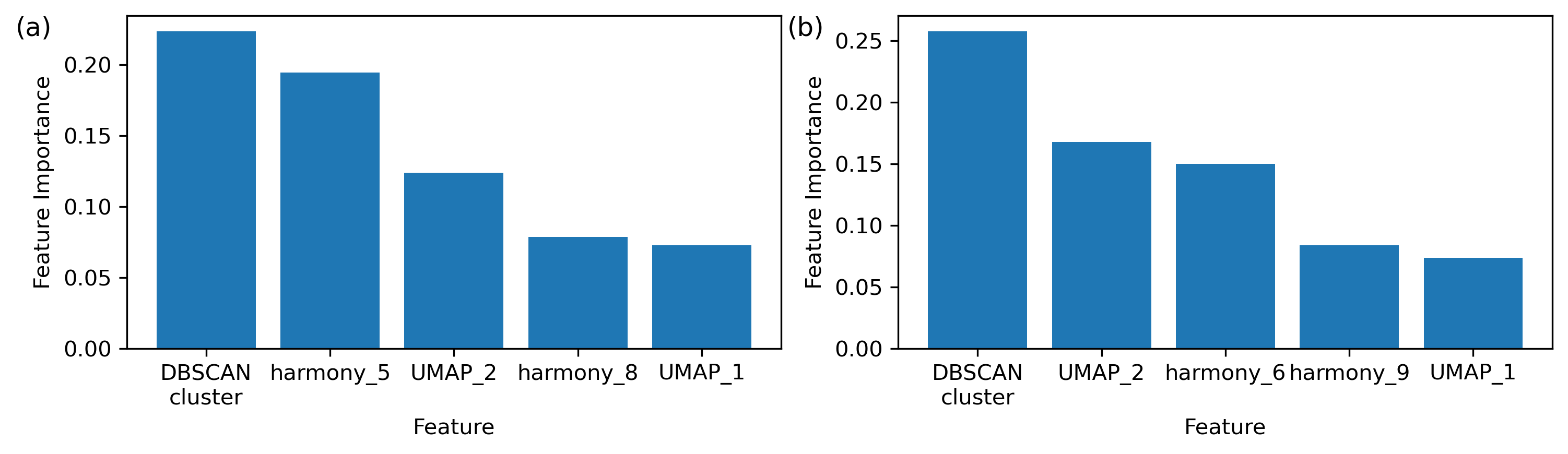}}
\caption{Top five features with the highest importance in CatBoost for Experiment 3, accounting for 70\% of total feature importance. These features span all three levels of resolution in both cases, with no single level dominating, indicating the classifier's effective utilization of diverse the resolutions of the features.}
\label{fig:featureImportance}
\end{figure}

\paragraph{\textbf{Impact of Multi-Resolution Feature Space on Model Performance}}
The construction of a multi-resolution feature space is a unique and innovative aspect of HiCat. Its supervised learning component, CatBoost, uses numerous shallow trees (parsimonious models) that sequentially address misclassifications from prior ensemble models. Each tree is expected to automatically select important features from the multi-resolution feature space tailored to specific misclassifications. Figure~\ref{fig:featureImportance} shows the top five features with the highest importance in two data analyses from experiment 3, demonstrating that features across all resolutions are selected by many shallow trees and play an essential role in HiCat’s predictions.

\section{Discussion}
We benchmarked our method against other leading approaches using 10 publicly available genomic datasets. The studies yielded three key findings: (1) Among methods tested on query data without new cell types, all performed well, but HiCat consistently delivered superior accuracy and cell type separation in most data pairs. (2) When query sets included unknown cell types, the performance of all methods dropped; however, HiCat remained the most resilient, achieving notably higher accuracy than its competitors. (3) To our knowledge, other methods do not directly tackle the challenge of distinguishing multiple unknown cell types. Conversely, HiCat excelled in these situations, effectively identifying and annotating unseen cell types, even with a limited number of cells in the query data. This exceptional performance stems from three design advantages of our method, which are discussed below.

\paragraph{\textbf{Integration of Reference and Query Information}}: A key advantage of HiCat is its ability to integrate genomic data from both reference and query sets, creating a unified embedding space. This hybrid approach offers two distinct benefits. First, by incorporating query samples into the feature engineering process, the effective sample size is significantly increased, allowing the model to learn from a richer set of patterns. Second, training the model on a unified embedding space makes it more transferable, enabling it to better classify samples in the query set. This combination of leveraging both reference and query data empowers HiCat to excel in scenarios where other methods struggle to identify meaningful patterns, especially when the query data includes new cell types.

\paragraph{\textbf{Sequential Dimensionality Reduction Steps and Multi-Resolution Feature Space}}: Another significant advantage of our approach is the use of multiple dimensionality reduction techniques, such as PCA and UMAP, followed by DBSCAN clustering. This sequential application of methods concentrates the information in high-dimensional data while retaining critical biological information. While 50-dimensional embeddings from PCA allow for more nuanced distinctions, the 2D UMAP embeddings provide a comprehensive overview of the data, and the clustering label is 1D interpretable dimension reduction directly related to cell types. Our model can effectively capture global and local data structures by establishing a 53-dimensional feature space that includes outputs from various resolution levels. The catBoost classifier is an ensemble learning from many shallow trees, where each tree is a parsimonious classification model that `automatically' picks important features from 53 candidates at different resolutions. This multi-resolution strategy enables HiCat to identify subtle patterns that other models (with fixed resolution of features) may overlook, particularly in complex cell types or mixed populations.

\paragraph{\textbf{Fusing Supervised and Unsupervised Signals}}: HiCat stands out by harnessing the advantages of both supervised and unsupervised learning. The supervised CatBoost model is trained with labeled reference data, ensuring accurate identification of known cell types, while unsupervised clustering using DBSCAN effectively identifies unknown cell types. This combined embedding approach creates a more robust feature space that can manage situations with new or sparse cell populations. By integrating these signals, HiCat circumvents the typical drawbacks of solely clustering-based methods, which often face issues with cluster impurity, and purely supervised methods, which struggle to incorporate novel cell types.

\paragraph{\textbf{Limitations}} HiCat performs well across various genomic data scenarios, yet certain limitations should be noted, highlighting opportunities for enhancement and exploration. Following are three examples. Firstly, our method assumes that the reference and query sets contain enough cells in shared cell types, ensuring significant overlap between the two datasets. This facilitates effective alignment during batch effect removal since the Harmony algorithm utilizes shared cell types to align both reference and query data in a unified space. However, when the reference set holds cell types with numerous samples absent from the query set, Harmony's performance may decline. Insufficient overlap can hinder its ability to align the datasets accurately, leading to suboptimal embeddings. Future research could investigate alternative batch effect removal methods, such as Seurat’s CCA (Canonical Correlation Analysis) or MNN (Mutual Nearest Neighbors) \cite{haghverdi_batch_2018,zhang_novel_2019,yang_smnn_2020}, which might perform better in scenarios with limited overlap. Secondly, each element of HiCat—like Harmony, UMAP, DBSCAN, and CatBoost—has been employed as standard methods with default settings in this research. Although this study aimed to showcase the pipeline's overall capabilities in managing unseen cell types, tuning the parameters of each component could notably enhance performance. For instance, revising the number of UMAP neighbors or adjusting DBSCAN clustering parameters might yield more precise clustering and improved differentiation of novel cell types. Furthermore, optimizing the CatBoost model's parameters, such as adjusting the learning rate or depth, could further boost classification accuracy. A major challenge in seeking optimal settings is the computational resources and datasets needed to thoroughly evaluate every possible parameter combination for each component within the HiCat pipeline. A comprehensive search for ideal configurations requires an extensive range of datasets across various conditions, which can be resource-intensive. Future investigations could prioritize hyperparameter optimization techniques, like grid search or Bayesian optimization \cite{klein_fast_2017}, to systematically pinpoint the most effective parameter combinations. Alternatively, creating a more adaptive pipeline that can adjust settings dynamically based on data characteristics presents an intriguing avenue for future research. Thirdly, the primary aim of this conference paper is to present our methodology. In a subsequent journal paper, where there will be fewer constraints on word count and time, we plan to conduct a more comprehensive evaluation of HiCat's performance by (1) including a broader range of available approaches and (2) discussing scenarios where the reference and query sets are generated from different platforms or species. The methods compared with HiCat in this work are selected based on their frequent use in comparisons with other published techniques, although newer and potentially better approaches exist. The 33 data pairs used in this study are designed to examine the cross-platform and cross-species performance of annotation methods, but here we have pooled all of them together for analysis. A future journal publication could delve into subgroup analyses to explore these factors in more detail.

\section{Conclusion}
HiCat presents a transformative solution for cell-type annotation by fusing the strengths of supervised and unsupervised learning, integrating genomic data from both reference and query sets, and utilizing a multi-resolution embedding space. Its adaptability, resilience, and capability to manage new cell types, even in limited populations, position it as a significant asset for genomic data analysis. Benchmark results convincingly show that HiCat surpasses current methods across multiple scenarios, especially when query datasets include new cell types, thereby highlighting its importance in intricate biological research.

\subsection*{Acknowledgement}
This work is supported by NSERC Discovery \# RGPIN-2021-03530 (LX), the Canada Research Chair \#CRC-2021-00232 (XZ), Michael Smith Health Research BC Scholar: \# SCH-2022-2553(XZ) and National Research Council Canada's Grant DHGA-121 (XZ, LX). This research was enabled in part by computational resource support provided by Westgrid (https://www.westgrid.ca) and the Digital Research Alliance of Canada (https://alliancecan.ca).

\subsection*{Statement of Contributions}
Study conceptualization, funding acquisition, and student mentorship (XZ, LX); methodology and experiment design (XZ, CB, LX); data preparation (KB); computer experiment (CB, KB); initial manuscript (CB). All authors contributed to revising the manuscript and have approved the final version.

%
%
%
\bibliographystyle{splncs04}
\bibliography{references}
%





\end{document}